\documentclass[prl,twocolumn,superscriptaddress]{revtex4}
\usepackage[colorlinks=True,linkcolor=blue,citecolor=blue,filecolor=blue,urlcolor=blue]{hyperref}

\date{\today}

\usepackage{natbib}
\usepackage{graphicx}
\usepackage{todonotes}
\usepackage{amssymb}
\newcommand{\fegete}{Fe$_3$GeTe$_2$}
\usepackage{xcolor}

\begin{document}


\title{Non-reciprocal magnons in a two dimensional crystal with off-plane magnetization}

\author{Marcio Costa}
\affiliation{Instituto de F\'isica, Universidade Federal Fluminense, 24210-346 Niter\'oi, RJ, Brazil}
\author{J. Fern\'andez-Rossier}
\altaffiliation{On leave from Departamento de F\'{i}sica Aplicada, Universidad de Alicante, 03690 San Vicente del Raspeig, Spain.}
\affiliation{QuantaLab, International Iberian Nanotechnology Laboratory, 4715-330 Braga, Portugal}
\author{N. M. R. Peres}
\affiliation{QuantaLab, International Iberian Nanotechnology Laboratory, 4715-330 Braga, Portugal}
\affiliation{Centro de F\'{i}sica das Universidades do Minho e Porto and Departamento de F\'{i}sica and QuantaLab, Universidade do Minho, Campus de Gualtar, 4710-057 Braga, Portugal}
\author{A. T. Costa}
\affiliation{QuantaLab, International Iberian Nanotechnology Laboratory, 4715-330 Braga, Portugal}

\begin{abstract}

Non reciprocal spin waves have a chiral asymmetry so that their energy is different 
for two opposite wave vectors. They are found in atomically thin ferromagnetic overlayers 
with in plane magnetization and are linked to the anti-symmetric Dzyaloshinskii-Moriya surface 
exchange.  We use an  itinerant fermion theory based on first principles calculations to predict 
that non-reciprocal magnons can occur in \fegete, the first stand alone metallic two dimensional 
crystal with off-plane magnetization. We find that both the energy and lifetime of magnons are 
non-reciprocal and we predict that acoustic magnons can have lifetimes up to hundreds of 
picoseconds, orders of magnitude larger than in other conducting magnets.
\end{abstract}

\maketitle

A defining property of elementary excitations in crystals, such as electrons, excitons, 
phonons, plasmons and magnons is their dispersion curve $E(\vec{q})$. In most cases, 
the dispersion curves satisfy the reciprocity relation $E(\vec{q})=E(-\vec{q})$, reflecting 
the equivalence between the excitation and its mirror image, i.e., their non-chiral nature.  
In condensed matter systems, non-reciprocal  energy dispersions occur under specific circumstances 
and elicit great attention. Examples are chiral~\cite{Haldane1988} and helical~\cite{KaneMele2005} 
edge states of topological phases of various excitations, including electrons, photons and magnons, 
as well as Rashba split bands in crystals lacking inversion symmetry and having strong 
spin orbit coupling~\cite{Xiao2012}.

A major driving force for chiral phenomena in magnetism~\cite{Herve2018,Han2019,Ding2020} is the 
antisymmetric exchange $\vec{D}_{ij}\cdot(\vec{S}_i\times\vec{S}_j)$, proposed by 
Dzyaloshinskii~\cite{Dzyaloshinskii1957} and Moriya~\cite{Moriya1960} (DM). This special type 
of super-exchange is enabled by the combination of spin orbit coupling~\cite{Moriya1960} and 
the absence of an inversion center between spins $i,j$. These conditions are naturally found 
in overlayers of atomically thin ferromagnets on top of surfaces with high spin orbit coupling. 
With this background, the existence of non reciprocal spin waves was 
predicted~\cite{Udvardi2009,Costa2010:SOCMethod}, provided that the DM vector $\vec{D}$ is parallel 
to the magnetization $\vec{M}$. Symmetry considerations for this class of systems~\cite{Crepieux1998} 
leads to the conclusion that the interfacial $\vec{D}$ lies in-plane, so that non-reciprocal 
spin waves in interfaces can only exist for ferromagnets with in plane easy axis, 
consistent with experimental observations~\cite{Zakeri2010,Zakeri2012}. Logical devices based on 
non-reciprocal spin-waves have been recently proposed~\cite{Jamali2013}.

In this work we show that non-reciprocal spin waves can exist in a newly discovered class of 
2D magnets~\cite{2d_magnets_rev}, stand alone two dimensional crystals with off-plane magnetization. 
The survival of magnetism in 2D is definitely linked to a strong spin orbit coupling, that opens up 
a gap in the magnon spectrum, preventing the infrared catastrophe that destroys long range 
order in isotropic 2D magnets, as shown by Mermin and Wagner\cite{MerminWagner1966}, 
inspired~\cite{Halperin2019} by Hohenberg~\cite{Hohenberg1967}. 

Here we explore magnons of \fegete\ for several reasons. First, it has a low symmetry magnetic unit cell, 
without an inversion center. Second, the observation of large Anomalous Hall effect~\cite{AHE}, 
anomalous Nerst effect~\cite{Xu2019} and skyrmions~\cite{skyrmions_fegete} in thin films strongly 
suggests that intrinsic DM interaction, as opposed to interfacial, is active in \fegete. Third, the 
system is a conductor, unlike other widely studied 2D crystals such as CrI$_3$, and has a large 
Curie temperature, that can reach room temperature upon gating~\cite{gate_fegete}.  
\fegete\ was synthesized for the first time, in  bulk, 
in 2006~\cite{fe3gete2FirstSynthesis}. Only much more recently, however, high quality few 
layers samples have been produced~\cite{Liu2017_MBE}. Monolayers have been obtained by 
exfoliation~\cite{Fei2018}.

Because of its conducting nature and high-temperature ordering, \fegete\ is closer to technological 
applications. On the theory side, modelling magnons in conducting ferromagnets represents a big 
challenge due to the non-integer nature of the magnetic moments, the long-range exchange, and the 
damping of magnons due to their coupling to Stoner excitations. A microscopic description that does 
not take the itinerant character into account, such as that provide by spin models, will fail to 
describe most of the relevant physics of these systems. 

\begin{figure}
    \centering
    \includegraphics[width=\columnwidth]{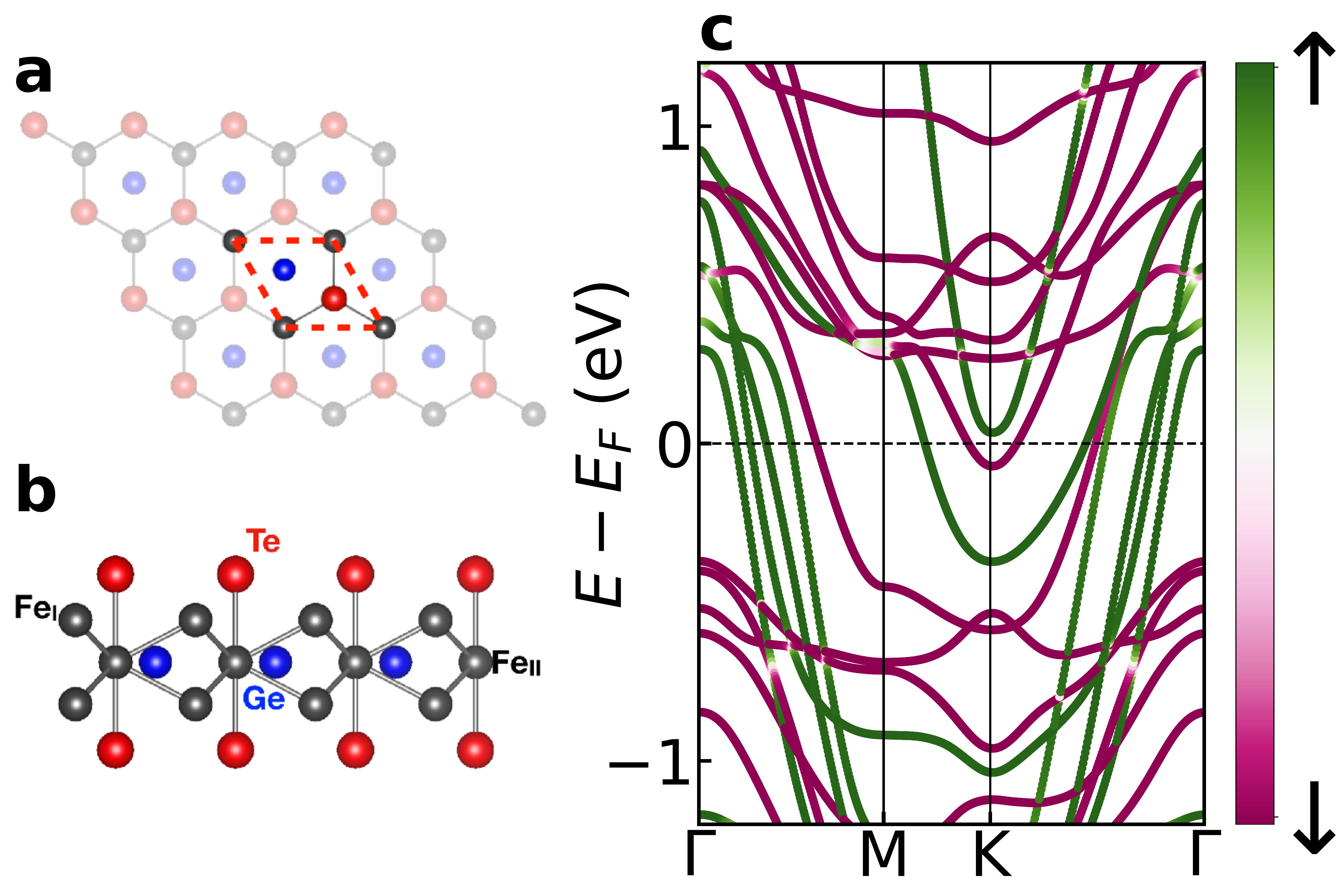}
    \caption{Top (\textbf{a}) and side (\textbf{b}) views of the lattice structure, 
showing the unit cell (marked by the dashed red line in \textbf{a}) and the two
nonequivalent Fe sites in (\textbf{b}). In (\textbf{c}) we show the band structure 
along high symmetry points in the 2D Brillouin zone (depicted in the inset of fig.~\ref{fig:dispersionGK}\textbf{a}). 
The color code shows the projection 
of the electronic eigenvectors on the eigenstates of $S_z$.}
    \label{fig:bands}
\end{figure}

We compute the magnon spectra of a \fegete\ monolayer using the itinerant fermion 
picture~\cite{Costa2010:SOCMethod,Costa2020:CrI3}. With this method we are able to 
extract magnon energies and lifetimes from a first principles electronic structure 
calculation, without the need of building an intervening effective spin model. 
We use Density Functional Theory (DFT)~\cite{Hohenberg64,Kohn65} to derive
an effective fermionic Hamiltonian to describe the spin dynamics of 2D materials.
The unit cell of \fegete is shown in Fig.~\ref{fig:bands}.  It has three Fe atoms, 
occupying two nonequivalent positions, $A$ and $B$. We denote them Fe$^A$, Fe$^{B_1}$ 
and Fe$^{B_1}$. There is no inversion center along the lines 
joining Fe$^{A}$ and Fe$^{B_{1,2}}$.

The DFT calculations were performed using the plane waves code \textsc{Quantum Espresso}~\cite{QE-2017}. 
The electronic exchange-correlation is described by the generalized gradient approximation (GGA) within 
the Perdew-Burke-Ernzerhof (PBE) functional~\cite{pbe}. Ionic cores are described using projector 
augmented wave (PAW) pseudopotentials~\cite{PAW}. The local effective paramagnetic Hamiltonian is 
obtained using a direct projection of the Kohn-Sham states onto pseudo-atomic orbital (PAO) 
basis~\cite{PAO2}, as implemented in the \textsc{Paoflow} code~\cite{PAO5}. 

The PAO tight-binding Hamiltonian is constructed using a $spd$ basis for Fe, Ge and Te atoms. 
We then add local spin-orbit coupling and intra-atomic Coulomb repulsion~\cite{costa2019}. 
The spin-orbit coupling strengths of Fe, Ge and Te are $\lambda_{\rm Fe}=50$~meV, 
$\lambda_{\rm Ge}=200$~meV, $\lambda_{\rm Te}=600$~meV~\cite{handbookphotochem}. 
The mean-field self-consistent ground state is obtained~\cite{costa2018_bismuthene,Costa2020:CrI3} 
by treating every component of the spin moment in each Fe atom as an independent variable.
The resulting band structure, shown in figure~\ref{fig:bands}, features several 
spin polarized bands at the Fermi energy, portraying \fegete\ as a ferromagnetic conductor. 

We find that the mean-field spin moments are $s_{B_1}=s_{B_2}=2.56 \mu_\mathrm{B}$ and 
$s_{A}=1.52 \mu_\mathrm{B}$, all of them along the off-plane axis. These values 
are in excellent agreement with the DFT results, $s_{B_1}=s_{B_2}=2.54 \mu_\mathrm{B}$ and 
$s_{A}=1.52 \mu_\mathrm{B}$. The spin moments of Te and Ge are negligible.  
We note that, given that the magnetic moments are approximately twice the spin values, 
the tentative spin values of Fe atoms are clearly not quantized as half integers.

\begin{figure}[t]
    \centering
    \includegraphics[width=0.8\columnwidth]{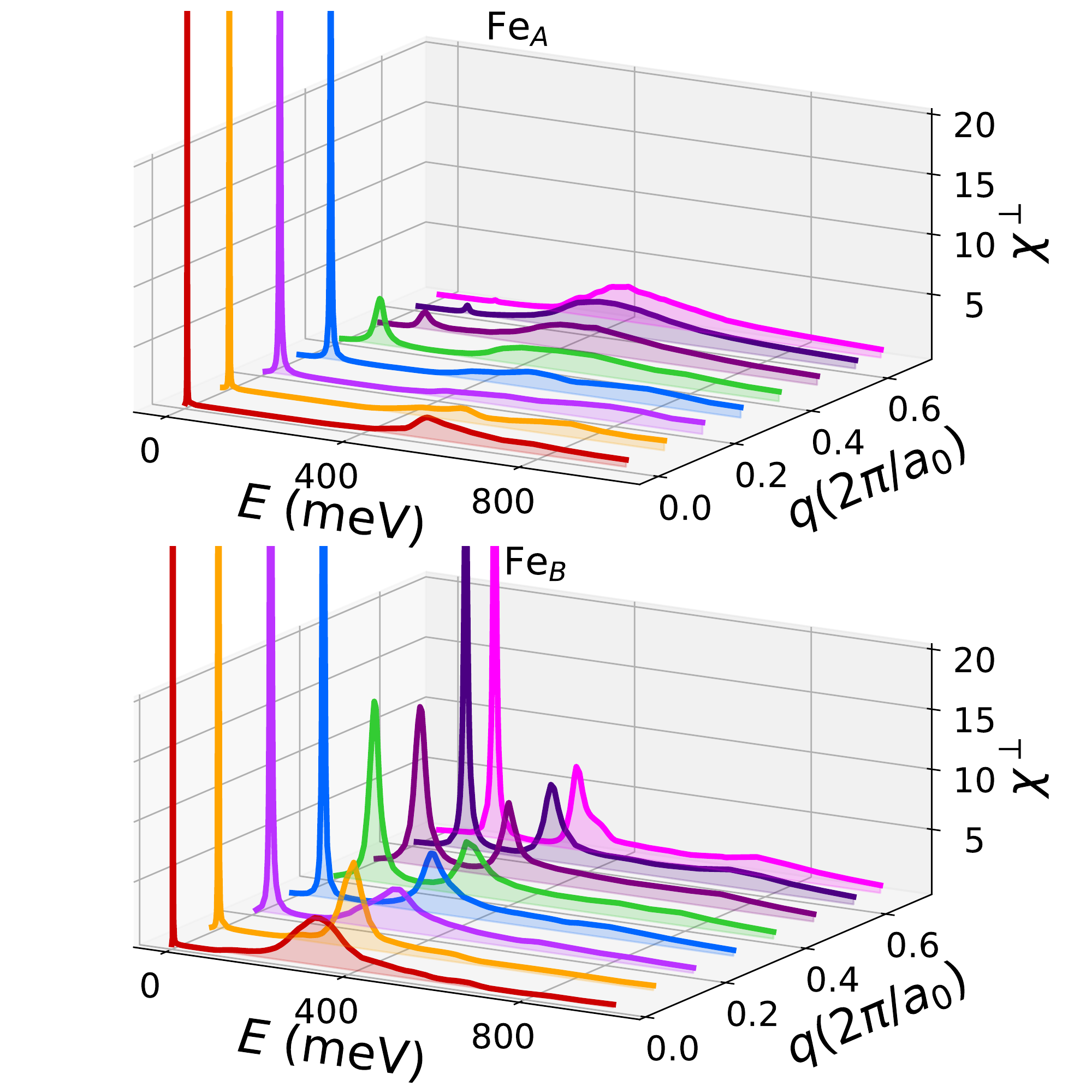}
    \caption{Magnon spectral density projected at the two nonequivalent Fe sites as a 
    function of energy, for a few selected wave vectors along the $\Gamma-K$ direction. The sharp peak at low energies is associated with the ``acoustic'' magnon and the broad structure at energies $\sim 300$~meV is the (strongly damped) ``non-bonding'' magnon.}
    \label{fig:ImchixqxE}
\end{figure}
\begin{figure}
    \centering
    \includegraphics[width=\columnwidth]{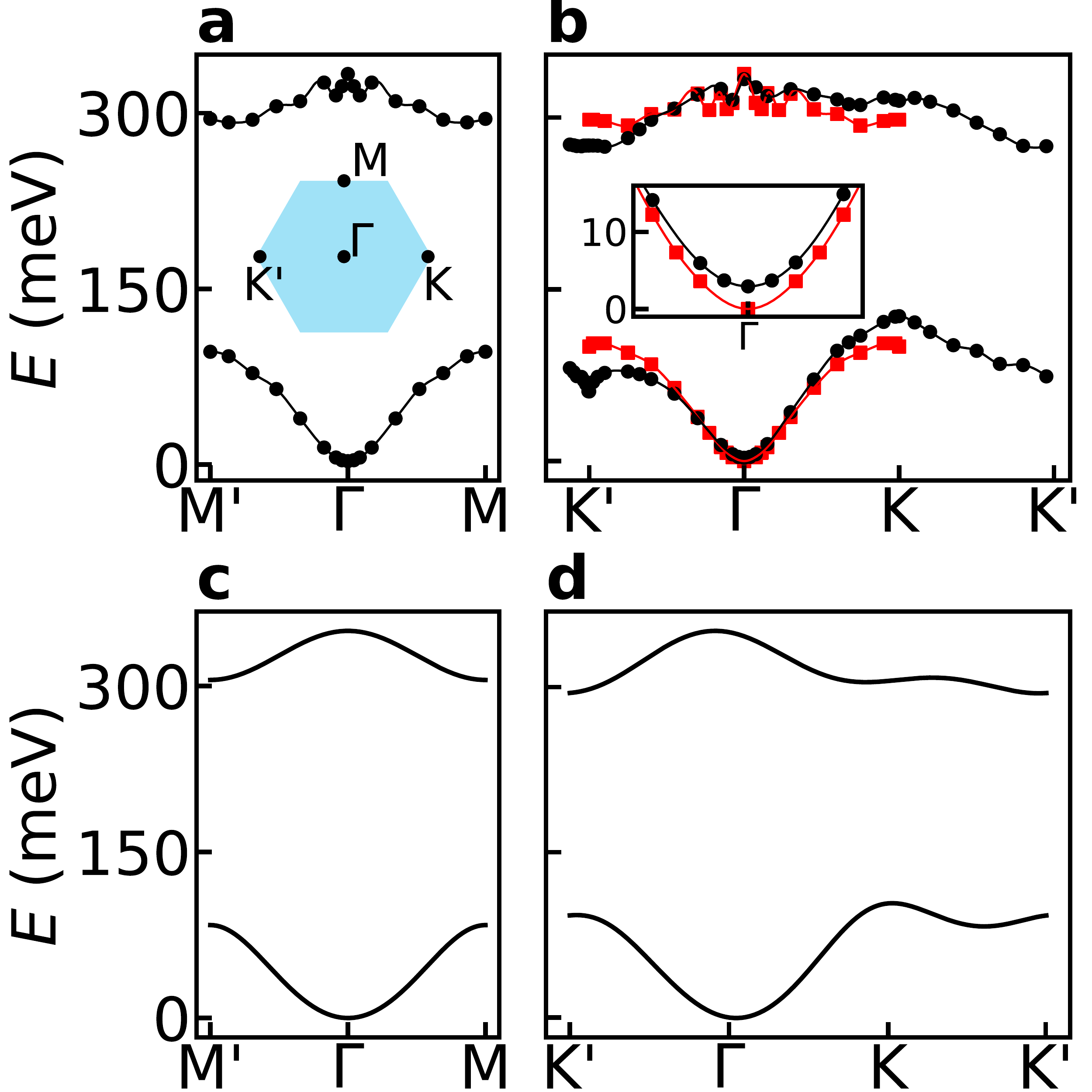}
    \caption{Fe$_{3}$GeTe$_{2}$ magnon dispersion relation along high 
             symmetry lines in the 2D Brillouin zone. The dispersion 
             relation along $\Gamma$-M (\textbf{a}) is reciprocal, whereas 
             along $\Gamma-K$ (\textbf{b}, black circles) it shows strong 
             non-reciprocity. In the inset we show a zoom of the dispersion relation 
             for the acoustic magnon close to the $\Gamma$ point, where the magnetocrystalline 
             anisotropy gap $\Delta_\Gamma\equiv E(\Gamma)=2.93$~meV can be clearly seen. 
             For comparison, we also show the dispersion relation calculated without spin-orbit 
             coupling (red squares), which is perfectly reciprocal and shows no anisotropy gap, 
             as expected. In (\textbf{c}) and (\textbf{d}) we show the dispersion relations obtained with the
             localized spins model, including the second neighbor Dzyaloshinskii-Moriya coupling.
The Brillouin zone is shown in the inset of panel \textbf{a}. }
    \label{fig:dispersionGK}
\end{figure}

The key quantity in the itinerant fermion theory for spin 
excitations~\cite{Costa2010:SOCMethod,Costa2020:CrI3} is the spin-flip spectral 
density ${\cal S}(E,\vec{q})\equiv{\rm Im}[\chi^\perp(E,\vec{q})]$, where
\begin{equation}
    \chi_{ll'}^\perp(E,\vec{q})\equiv\int_{-\infty}^\infty dt e^{-i\frac{E}{\hbar}t}
    \left\{-i\theta(t)\left\langle[S^+_{l,\vec{q}}(t),S_{l',-\vec{q}}^-(0)]\right\rangle\right\},
    \label{chi}
\end{equation}
$l,l'$ are atomic site indices, $E$ is the excitation energy, $\vec{q}$ is the magnon wave vector, 
$\theta(t)$ is the Heaviside unit step function and $\langle\cdot\rangle$ denotes thermal average. 
The four fermion correlator in eq.~(\ref{chi}) is computed in the 
Random Phase Approximation~\cite{Costa2006IntermQ,Costa2010:SOCMethod,Costa2020:CrI3}.

The diagonal entries, $\chi_{AA}^\perp(E,\vec{q})$ and 
$\chi_{B_1B_1}^\perp(E,\vec{q})=\chi_{B2B2}^\perp(E,\vec{q})$ of the spin-flip
spectral density are shown in figure~\ref{fig:ImchixqxE} for a few 
selected wave vectors. For a given value of $\vec{q}$ the spin flip spectral density has, 
in general, two types of features. First, symmetric peaks, with a width $\Delta E$ much smaller 
than peak energy $E$. These peaks are not present in the spectral density of the non-interacting 
susceptibility. These are magnons modes, featured by all ferromagnets. Second, broad asymmetric 
features, that correspond to the so called Stoner excitations and are only present in conducting 
ferromagnets.

Two well defined magnon branches are identified in Fig.~\ref{fig:ImchixqxE}. For reasons that will
become apparent later, we refer to the lower energy, narrow peaks as the acoustic branch and to 
the higher energy, broader peaks as the non-bonding branch. When SOC is included, the acoustic 
branch has a gap at the $\Gamma$ point is $\Delta_\Gamma = 2.9$~meV, that accounts for the magnetic 
anisotropy. Its magnitude is compatible with existing measurements~\cite{neutronscattering} and DFT 
calculations~\cite{Zhuang2016}. The acoustic branch has weight distributed between $A$ and $B$ 
sublattices, although most of it lies on the $B$ sites. In contrast, the non-bonding branch is 
missing entirely from the $A$ site. A broad feature appears at higher ($\gtrsim 400$~meV), energies, 
localized in the $A$ site, whose nature is discussed below.

The magnon dispersion relation along high-symmetry lines in the Brillouin zone is shown 
in figure~\ref{fig:dispersionGK}, calculated both with and without spin orbit coupling.
The bandwidth of the acoustic magnon ($\sim 120$~meV) is much larger than that obtained 
in other 2D magnets, such as CrI$_3$~\cite{Costa2020:CrI3}, and reflects a large exchange 
coupling between the magnetic moments in neighbouring Fe atoms, in line with the larger 
Curie temperature of \fegete . 

Importantly, when the SOC is included in the calculation, 
both the acoustic and the non-bonding bands become non-reciprocal in the $K-\Gamma-K'$ 
direction, but not on the $\Gamma-M$ direction. It is noteworthy that the 
dispersion relation of the acoustic mode around the $\Gamma$-point fits almost perfectly 
to a function of wave vector $q$ of the form $\Delta_\Gamma + Dq^2$, with negligible linear 
component. This is in contrast with the behavior of magnons in ultrathin transition metal 
films on heavy substrates~\cite{Costa2010:SOCMethod}, where a sizeable linear term is 
induced by the DM coupling, and has also been observed in relation
to the calculation of static spin spirals in \fegete~\cite{Manchon2020}.

At this point we introduce a model Hamiltonian for the magnons, in order to gain physical 
insight on the origin of the most salient features of the results obtained with the 
itinerant model. The departure point is a spin Hamiltonian
\begin{equation}
{\cal H}= {\cal H}_{\rm Heis}+ {\cal H}_{\rm DM}+{\cal H}_{\rm anis}, 
\label{spinhamiltonian}
\end{equation}
composed of an isotropic Heisenberg term ${\cal H}_\mathrm{Heis}$,
a Dzyaloshinskii-Moriya interaction ${\cal H}_\mathrm{DM}$ and a single-ion
anisotropy term ${\cal H}_\mathrm{anis}$. Explicit expressions and further detail
can be found in the supplemental material (SM). We build the magnon model using the 
conventional Holstein-Primakoff linear spin wave theory for a quantized spin model. 
The spins live in a \textsl{decorated honeycomb} lattice with three sites per unit cell, 
$A,B_1,B_2$ (see Fig.~1 of the SM), with spins $S_A$ and $S_B$.

Given that sites $B_1$ and $B_2$ are equivalent, we can introduce two new modes, symmetric 
and anti-symmetric combinations of $B_1$ and $B_2$, so that one of them becomes effectively 
decoupled from $A$. The decoupling naturally leads to three bands. One is associated with the
anti-symmetric $B$ mode. The other two describe a honeycomb ferromagnet with broken 
inversion symmetry, on account of the different nature of $A$ and $B$, and are separated 
by a gap. The projections of the magnon wave functions over the different sites 
(see Fig.~3 in the SM) show that the spin model naturally accounts 
for the fact that the acoustic branch is predominantly located in 
the symmetric $B$ mode, the $\vec{k}$ dependence of the weight on the $A$ site, and the 
complete localization of non-dispersive band on the $B$ mode. This behavior
is qualitatively identical to that of the magnon wave functions extracted directly from
the fermionic model (see SM for details).

We are now in position to address the origin of the non-reciprocal dispersion, obtained 
with the itinerant model, using the spin model. The fact that it only arises when spin 
orbit coupling is included is a clear indication that the its origin has to come from the 
non-Heisenberg terms in the Hamiltonian. We have considered both first and second neighbour 
DM couplings, $D_{a,a'}^{(1)}$ and $D_{a,a'}^{(2)}$, where $a,a'$  label the sites in the 
unit cell that do not possess an inversion center. We only consider the DM vector $\vec{D}$
parallel to the magnetization, \textit{i.e.}, in the off-plane direction.
 
We find that first and second neighbour DM coupling yield non-reciprocal dispersions. However,  
only a finite $D^{(2)}$ coupling for the B sublattice gives a non-reciprocal dispersion in the 
$K-\Gamma-K'$ line, and reciprocal dispersion along the $\Gamma-M$ direction. Therefore,
the non-reciprocal dispersion is consistent with a second neighbour DM interaction in the 
$B$ sublattice, for which the super-exchange pathways occurs via Tellurium atoms, the ones with 
the largest SOC in the crystal.

\begin{figure}
\includegraphics[width=\columnwidth]{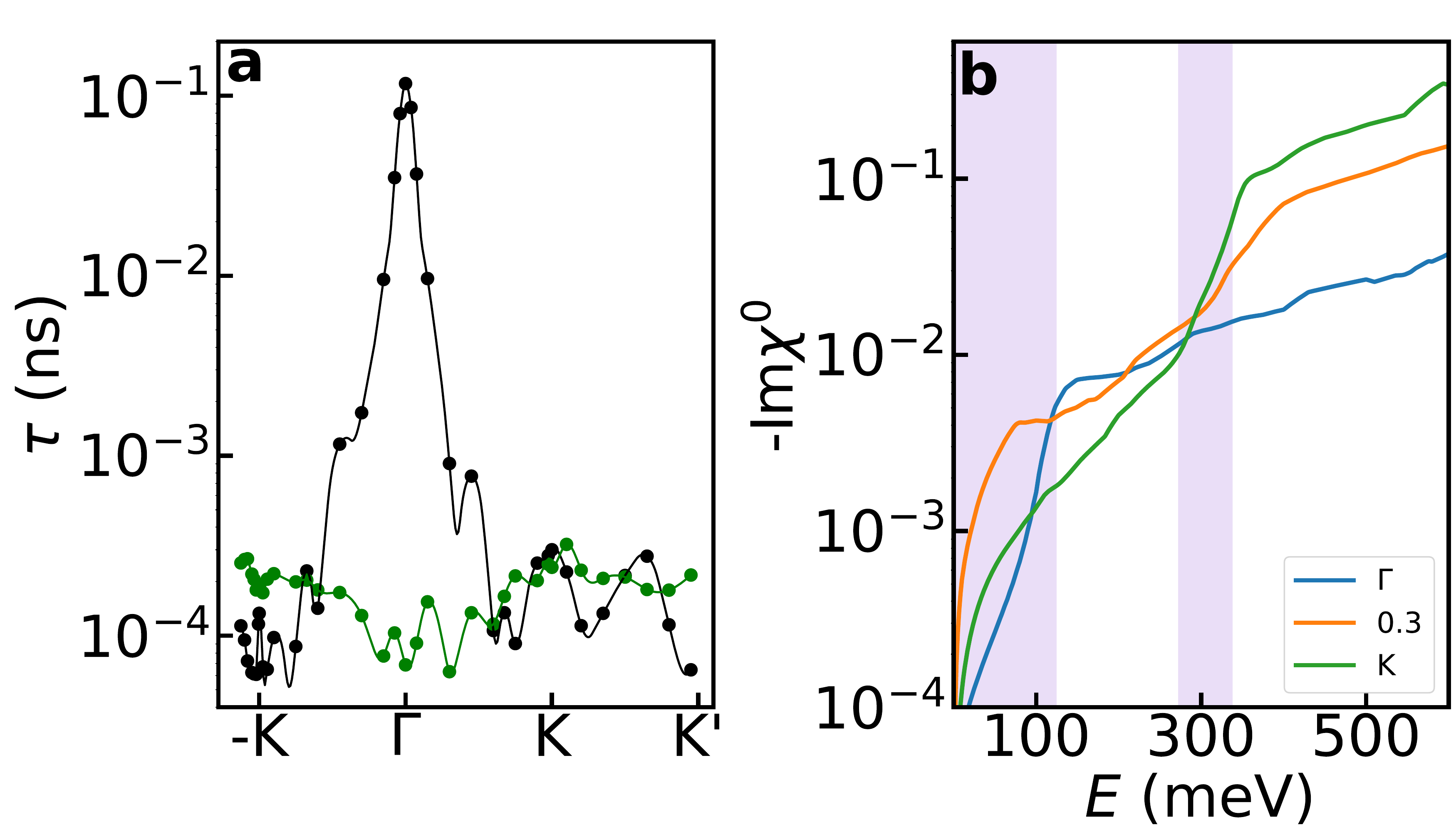}
\caption{\textbf{a}: Magnon lifetimes as a function of wave vector. \textbf{b}: Spectral density
of Stoner modes as a function of energy for three different wave vectors along the $\Gamma-K$ line.
The shaded regions mark the bandwidths of the acoustic and non-bonding magnons.}
\label{fig:stonerandtau}
\end{figure}

We now shift our attention to one of the hallmarks of itinerant magnetism: the fact that magnons 
have finite lifetimes, because of their coupling with the continuum of uncorrelated electron-hole 
excitations known as the Stoner continuum. In figure~\ref{fig:stonerandtau}\textbf{a} we show the 
magnon lifetimes as a function of wave vector. The lifetime is related to the linewidth of the
spectral density via $\tau\equiv \frac{2\hbar}{\Delta E}$. Remarkably, the acoustic magnons close 
to the $\Gamma$ point have very long lifetimes ($\sim 100$~ps), given that the longest lifetimes 
measured in ultrathin conducting magnets~\cite{Qin2015} are $\sim 0.4$~ps. A long magnon lifetime is a 
very important figure of merit for potential applications of magnons as carriers of information, 
for example. 

The lifetime of a magnon with energy $E$ and wave vector $\vec{q}$ scales 
inversely with the weight of the Stoner spectral density at the same energy and wave vector.
Due to the spin polarization of the $d$ bands, the density of Stoner modes 
is very small for energies much smaller than the exchange splitting 
(roughly proportional to magnetization and the intra-atomic Coulomb repulsion strength).
It grows abruptly as the excitation energy approaches the exchange splitting, 
as seen in fig.~\ref{fig:stonerandtau}\textbf{b}. For \fegete , the energies of acoustic magnons lie 
in the region of small density of Stoner modes, whereas the non-bonding magnons live
in the energy range where the Stoner spectral density is considerable. This is the origin
of the large difference between acoustic and non-bonding magnons lifetimes.

In this context, we can understand why the itinerant picture leads to only 
two magnon modes, whereas the localized spins model has three. Basically, the
third magnon band, still higher in energy than the second, is degenerate with the 
continuum of Stoner spin flip excitations. As a result,  
the spectral weight of the high energy optical magnon mode is transferred to the incoherent 
features predominantly localized in the $A$ site, shown in Fig.~\ref{fig:ImchixqxE}. 
The difference between the two theories highlights the limitations of the spin Hamiltonian,  
most notably in the case of itinerant magnets.

The acoustic magnon lifetimes are also non-reciprocal. This effect is \textsl{not} 
exclusively related to the the non-reciprocity of the energy dispersion: 
lifetimes are shorter in general for higher energy states. We find that, although 
magnons around the K point have both energies and lifetimes larger than those at $K'$.
The ultimate reason of this non-reciprocal lifetimes stems from the fact that the 
density of Stoner modes in \fegete\ is also non-reciprocal.
  
Besides endowing magnons with finite lifetimes, the Stoner continuum renormalizes the 
magnon energies, much like friction changes the natural frequency of an harmonic oscillator.
This is the origin of the oscillations in the dispersion relation of the non-bonding magnons,
seen in figure~\ref{fig:dispersionGK}c. The dispersion relation of the acoustic magnons 
close to the K~point also display some oscillations of the same origin. 


In conclusion, we have calculated magnons in monolayer \fegete\ using an itinerant 
fermion description derived from first principles calculations, and we have compared 
those results with the simple magnon theory for a spin model Hamiltonian for a decorated 
honeycomb lattice with three spins per unit cell. Due to broken mirror symmetry and spin-orbit 
coupling, magnons' energies and lifetimes show non-reciprocal behavior along the 
$\Gamma-K$ direction. Our findings are consistent with a second neighbour DM coupling 
in the B sublattice, but this deserves further attention. The coupling of magnons to Stoner 
excitations results in a intrinsic broadening of the two lowest energy magnon branches, 
and the melting of the optical mode, expected in the spin model, into a broad spectral feature 
at high energies. From our results we infer a value for the exchange stiffness  
that is compatible with the large magnetic transition temperatures observed 
experimentally. Furthermore, we find that the acoustic magnons 
are extremely long-lived for a conducting two-dimensional 
ferromagnet ($\tau\sim 100$~ps at the $\Gamma$ point), which make this material potentially 
very useful for magnonics and spintronics applications. 
Our work shows that non-reciprocal magnons can exist in  2D crystals with off-plane magnetization due 
to their intrinsic DM interaction and suggest that \fegete\ is a very interesting material to 
explore non-trivial magnon effects.

\begin{acknowledgments}

N. M. R. P.  acknowledges support from the European Commission through the project "Graphene- Driven Revolutions in ICT and Beyond" (Ref. No. 881603 -- Core 3), and the Portuguese Foundation for Science and Technology (FCT) in the framework of the Strategic Financing UID/FIS/04650/2013,   COMPETE2020, PORTUGAL2020, FEDER and the Portuguese Foundation for Science and Technology (FCT) through projects PTDC/FIS-NAN/3668/2013 and POCI-01-0145-FEDER-028114. 
JFR  acknowledges financial support from FCT for UTAP-EXPL/NTec/0046/2017 projects, 
as well as Generalitat Valenciana funding Prometeo2017/139 and MINECO-Spain (Grant No. MAT2016-78625-C2)
ATC acknowledges the use of computer resources at MareNostrum and the technical support provided by 
Barcelona Supercomputing Center (RES-FI-2019-2-0034, RES-FI-2019-3-0019).

\end{acknowledgments}

\bibliographystyle{apsrev4-1}

\end{document}



\title{Supplemental Material for \\ Non-reciprocal magnons in a two dimensional crystal with off-plane magnetization}

\author{Marcio Costa}
\affiliation{Instituto de F\'isica, Universidade Federal Fluminense, 24210-346 Niter\'oi, RJ, Brazil}
\author{J. Fern\'andez-Rossier}
\altaffiliation{On leave from Departamento de F\'{i}sica Aplicada, Universidad de Alicante, 03690 San Vicente del Raspeig, Spain.}
\affiliation{QuantaLab, International Iberian Nanotechnology Laboratory, 4715-330 Braga, Portugal}
\author{N. M. R. Peres}
\affiliation{QuantaLab, International Iberian Nanotechnology Laboratory, 4715-330 Braga, Portugal}
\affiliation{Centro de F\'{i}sica das Universidades do Minho e Porto and Departamento de F\'{i}sica and QuantaLab, Universidade do Minho, Campus de Gualtar, 4710-057 Braga, Portugal}
\author{A. T. Costa}
\affiliation{QuantaLab, International Iberian Nanotechnology Laboratory, 4715-330 Braga, Portugal}

\maketitle

\section{The spin model}

\begin{figure}
\includegraphics[width=\columnwidth]{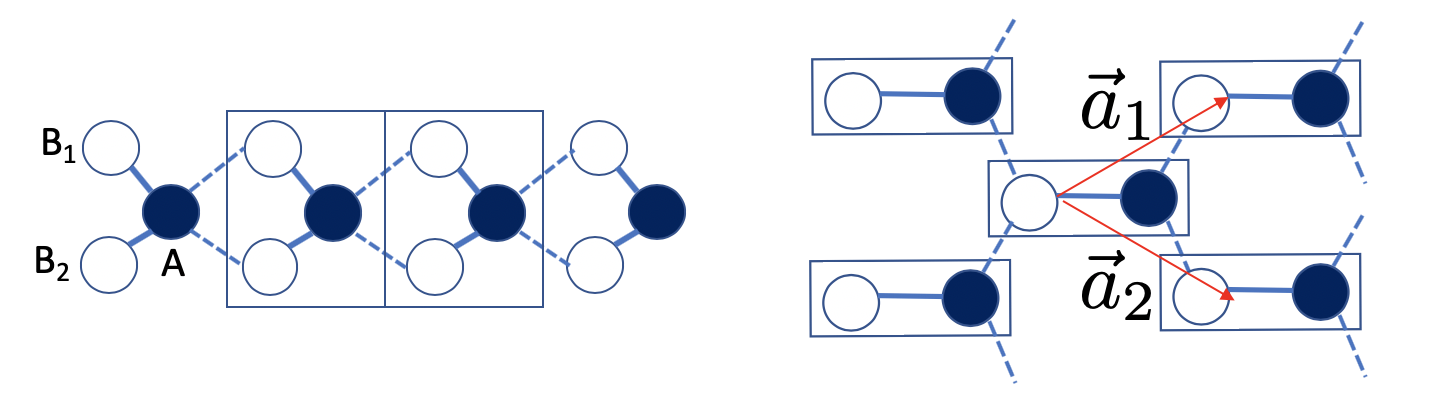}
\caption{Decorated Honeycomb lattice.  Left panel: side view, 
showing 3 atoms per unit cell. Right panel: Top view, showing 
the honyecomb arrangment, and the Brag vectors $\vec{a}_1$ and $\vec{a}_2$.}
\label{fig:honey}
\end{figure}

We derive here the magnon Hamiltonian for a quantum spin model defined on 
a decorated honeycomb lattice. 
We consider a unit cell with 3 atoms, $A$, $B_1$ and $B_2$, depicted in fig.~\ref{fig:honey}. 
Atoms $B_1$ and $B_2$ 
are stacked vertically on top of each other. When seen from the top, the decorated 
honeycomb lattice  looks like a regular honeycomb lattice, as atoms $B_1$ and $B_2$ 
are aligned. 

We shall consider a Hamiltonian with three types of coupling: Heisenberg, 
Dzyaloshinskii-Moriya (DM) and single ion anisotropy,
\begin{equation}
{\cal H}= {\cal H}_{\rm Heis}+ {\cal H}_{\rm DM}+{\cal H}_{\rm anis} 
\label{spinhamiltonian}
\end{equation}
In order to expedite the derivation of the spin wave theory taking advantage
of the crystal translational invariance, we write the spin Hamiltonian making explicit
the unit cell index $\vec{R}$ an the intra cell label $a=A,B_1,B_2$. The Heisenberg coupling reads,
\begin{equation}
 {\cal H}_{\rm Heis}=  \sum_{\vec{R},\vec{R}',a,a'} J_{a,a'}(\vec{R}-\vec{R}')\vec{S}(\vec{R},a)\cdot\vec{S}(\vec{R}',a') 
 \end{equation}
Below we consider up to second neighbor exchange. 

Taking into account only the $z$ component of the Dzyaloshinskii-Moriya vector, the DM term reads,
\begin{eqnarray}
 {\cal H}_{\rm DM}=  
- \sum_{\vec{R},\vec{R}', a,a'} D_{a,a'}(\vec{R}-\vec{R}')
\nonumber \\
\left[S_x(\vec{R},a)S_y(\vec{R}',a')
-S_y(\vec{R},a)S_x(\vec{R}',a')\right].
 \end{eqnarray}
The single ion anisotropy reads,
\begin{equation}
 {\cal H}_{\rm anis}=\sum_{a,R}  -|K_a| S_z^2(\vec{R},a) .
\end{equation}

\section{Linear spin wave theory}
We now introduce the standard bosonic Holstein-Primakoff (HP) representation 
of the spin operators~\cite{Holstein1940}. We assume that the magnetization of
the classical ground state lies along the $\hat{z}$ axis and we keep only quadratic 
terms in the boson operator, which amounts to ignoring magnon-magnon interactions. 
This is the so-called linear spin wave theory. This is accomplished by expressing 
the spin operators  in terms of the HP  bosons $c$ as
\begin{eqnarray}
\vec{S}(\vec{r}) \cdot\vec\Omega(\vec{r})&=&  S- c^{\dagger}(\vec{r})c(\vec{r})
\nonumber \\
S^{(+)}(\vec{r}) &\simeq& \sqrt{2S}  c (\vec{r})\nonumber \\
S^{(-)}(\vec{r}) &\simeq&  \sqrt{2S}   c^{\dagger}(\vec{r}),
\label{HPrep}
\end{eqnarray}
where $\vec{r}$ labels every spin in the crystal. Thus, $\vec{r}$ is completely 
specified by $\vec{R},a$.  The representation of the $S_z^2$ terms we only keep 
terms quadratic in the $c$ operators, dropping the for boson operators.

\subsection{Momentum representation of the magnon Hamiltonian}
After a lengthy but straightforward calculation, we  derive the a quadratic 
Hamiltonian for the HP bosons in the reciprocal space. For that matter, we 
represent the \ref{spinhamiltonian} using the \ref{HPrep}, keeping only terms 
bilinear in the HP bosons, and we define
\begin{equation}
c_{\vec{k},a}=\frac{1}{\sqrt{N}}\sum_{\vec{R}} c_{\vec{R},a} e^{-i\vec{k}\cdot\vec{R}}.
\label{Fou0}
\end{equation}
We consider Heisenberg and DM coupling between first and second neighbors. The 
resulting magnon Hamiltonian reads,
 \begin{eqnarray}
{\cal H}=
 \sum_{\vec{k},a,a'} c_{\vec{k},a}^{\dagger}{\cal H}_{a,a'}(\vec{k})  c_{\vec{k}a'},
\end{eqnarray}
with
 \begin{eqnarray}
 {\cal H}_{a,a'}(\vec{k})\equiv   {\cal H}^{\rm Heis}_{a,a'}
 + {\cal H}^{\rm DM}_{a,a'}
 +{\cal H}^{\rm anis}_{a,a'} .
\end{eqnarray}
The wave vector $\vec{k}$ spans the Brillouin zone of the triangular lattice.

\subsection{Decoupling transformation and analytical expressions}

Explicit expressions are given below. Importantly, the Hamiltonian has the the form
\begin{eqnarray}
{\cal H}_{a,a'}(\vec{k}) \equiv
 \left(
\begin{array}{ccc}
h_{AA}(\vec{k}) & h_{AB}(\vec{k}) & h_{AB}(\vec{k}\\
h_{AB}(\vec{k})^* & h_{BB(\vec{k})} & h_{12}(\vec{k})\\
h_{AB}(\vec{k})^*& h_{12} (\vec{k})& h_{BB}(\vec{k})
\end{array}
\right),
\end{eqnarray}
reflecting the equivalence of sites $B_1$ and $B_2$.    
 By introducing a change of basis for the $B_1$ and $B_2$ sites, that 
defines the symmetric and antisymmetric modes, 
$c_{\vec{k},\pm}=\frac{1}{\sqrt{2}}\left(c_{\vec{k},B1}\pm c_{\vec{k},B2}\right)$, 
we can write the magnon Hamiltonian matrix, in the $A,+,-$ basis as,
 \begin{eqnarray}
{\cal H}_{a,a'}(\vec{k}) \equiv
 \left(
\begin{array}{ccc}
h_{AA}& \sqrt{2}h_{AB} & 0\\
\sqrt{2}h_{AB}^* & h_{BB}+h_{12} & 0\\
0& 0 & h_{BB}-h_{12}
\end{array}
\right)
\label{deco}
\end{eqnarray}
where we have used that  $h_{12}$ is real and we have removed the explitit dependence on $\vec{k}$ 
to simplify the notation.  This equation shows that the antisymmetric ($-$) mode
is decoupled from the other two.

  The remaining coupled modes, $A$ and $B_+$,  are isomorphic to quasiparticles in a honeycomb lattice with broken inversion symmetry, on account of the different couplings and spins of $A$ and $B$ sites.
The decoupled expression (\ref{deco}) permits to obtain analytical expressions for the magnon bands, 
\begin{eqnarray}
E_{\rm ac}(\vec{k})= \epsilon_0(\vec{k})-\sqrt{\epsilon_1(\vec{k})^2+ 2 |h_{AB}(\vec{k})|^2}, \nonumber \\
E_{\rm op}(\vec{k})= \epsilon_0(\vec{k})+\sqrt{\epsilon_1(\vec{k})^2+ 2 |h_{AB}(\vec{k})|^2}, \nonumber \\
E_{\rm non-bonding}(\vec{k})=h_{BB}(\vec{k})-h_{12}(\vec{k}),
\label{anali}
\end{eqnarray}
where we have defined 
\begin{eqnarray}
\epsilon_0(\vec{k})= \frac{h_{AA}(\vec{k})+h_{BB}(\vec{k})+h_{12}(\vec{k})}{2},\nonumber\\
\epsilon_1(\vec{k})= \frac{h_{AA}(\vec{k})-h_{BB}(\vec{k})-h_{12}(\vec{k})}{2}.  
\end{eqnarray}
The acoustic and optical bands $E_{\rm ac}(\vec{k})$, $E_{\rm op}(\vec{k})$ are associated 
to magnons that live both in the $A$ and $B$ sublattice, with the $B$ modes in phase,  
whereas $E_{\rm non-bonding}(\vec{k})$ is {\em sublattice polarized}.

In the following we derive explicit expressions of the magnon bands in terms of the exchange, 
DM and anisotropy terms, and derive their values so that the simple model agrees with the 
magnon bands obtained with the itinerant fermion picture.

\subsection{Magnon bands for the isotropic case}

We consider first the case only symmetric exchange interactions are included.  This permits
to compare with the calculations without SOC.

We introduce the first neighbor exchange coupling constants,  $J^{(1)}_{AB}$ and $J^{(1)}_{BB}$ as well 
as the second neighbour exchange for the $A$ sublattice, $J^{(2)}_{AA}$,  and the
intralayer (or direct)  and interlayer (or crossed) for the $B$ sublattices
$J^{(2)}_{BB,2d}$, $J^{(2)}_{BB,2c}$.

\subsubsection{First neighbour approximation}

We now keep only the first neighbour interaction. 
Introducing the notation $\phi_n=\vec{k}\cdot\vec{a}_n$, with $n=1,2$, we can write
\begin{eqnarray}
h_{AB}(\vec{k}) &=&-\sqrt{S_A S_B} J^{(1)}_{AB} (1+ e^{i\phi_1}+e^{i\phi_2})\nonumber\\
h_{AA}&=&6 J^{(1)}_{AB} S_B \nonumber\\
h_{BB}&=&3 J^{(1)}_{AB}S_A+J^{(1)}_{BB} S_B \nonumber\\
h_{12}&=&-J^{(1)}_{BB}S_B
\end{eqnarray}
Here the factors $6$ and $3$ reflect the number of first neighbours of the $A$ and $B$ sublattice
that are connected via the $J^{(1)}_{AB}$.
We obtain:
 $$\epsilon_0(\vec{k})= \frac{3J^{(1)}_{AB}}{2}\left(2  S_B+ S_A\right) $$
 and
 $$\epsilon_1(\vec{k})=\frac{3J^{(1)}_{AB}}{2}(2  S_B- S_A).  
  $$

 We now obtain the magnon energies at the high symmetry points.
 At $\Gamma$ and $K$ we have  $f(\Gamma)=3$ and  $f(K)=f(K')=0$.  We thus have: 
 \begin{eqnarray}
&E_{\rm ac}(\Gamma)=0   & E_{\rm ac}(K)=3J^{(1)}_{AB}S_A \nonumber\\
&E_{\rm op}(\Gamma)=  3J^{(1)}_{AB}\left(2  S_B+ S_A\right)& E_{\rm op}(K)=6J^{(1)}_{AB}S_B \nonumber
\end{eqnarray}
whereas the non-bonding state is dispersion less: 
$$E_{\rm non-bonding}=E_{\rm ac}(K)+ 2 J^{(1)}_{BB} S_B. $$

 From these equations we see that the gap between the optical and acoustic branch is due both  to the different inter-sublattice coordination of $A$ (6) and $B$ (3) sites and their different spin $S_A\neq S_B$. 
We also see that the non-bonding state is inside the gap, as long as  $2 J^{(1)}_{BB} S_B<
3J^{(1)}_{AB}(2S_B-S_A)$.  

We now use these results to obtain a rough estimate of the first neighbour exchange constants. 
At this level of the theory, the values of $S_A$ and $S_B$  do not need to be integer or half-integer, and we take   we take $S_A= M_A/2 =0.76 $ and $S_B=M_B/2= 1.26$.       The value of the acoustic band at the $K$ point permits to estimate $J^{(1)}_{AB}$, using $3J^{(1)}_{AB}S_A\simeq 100$ meV. We obtain  $J^{(1)}_{AB}\simeq 44$ meV

Given that the band at around 300 meV is exclusively localized in the $B$ sublattice, we infer that this is the non-bonding band.  We thus  have the equation $2 J^{(1)}_{BB} S_B\simeq200$ meV, from which we obtain  $J^{(1)}_{BB}\simeq 80$ meV.

\begin{figure}
\includegraphics[width=\columnwidth]{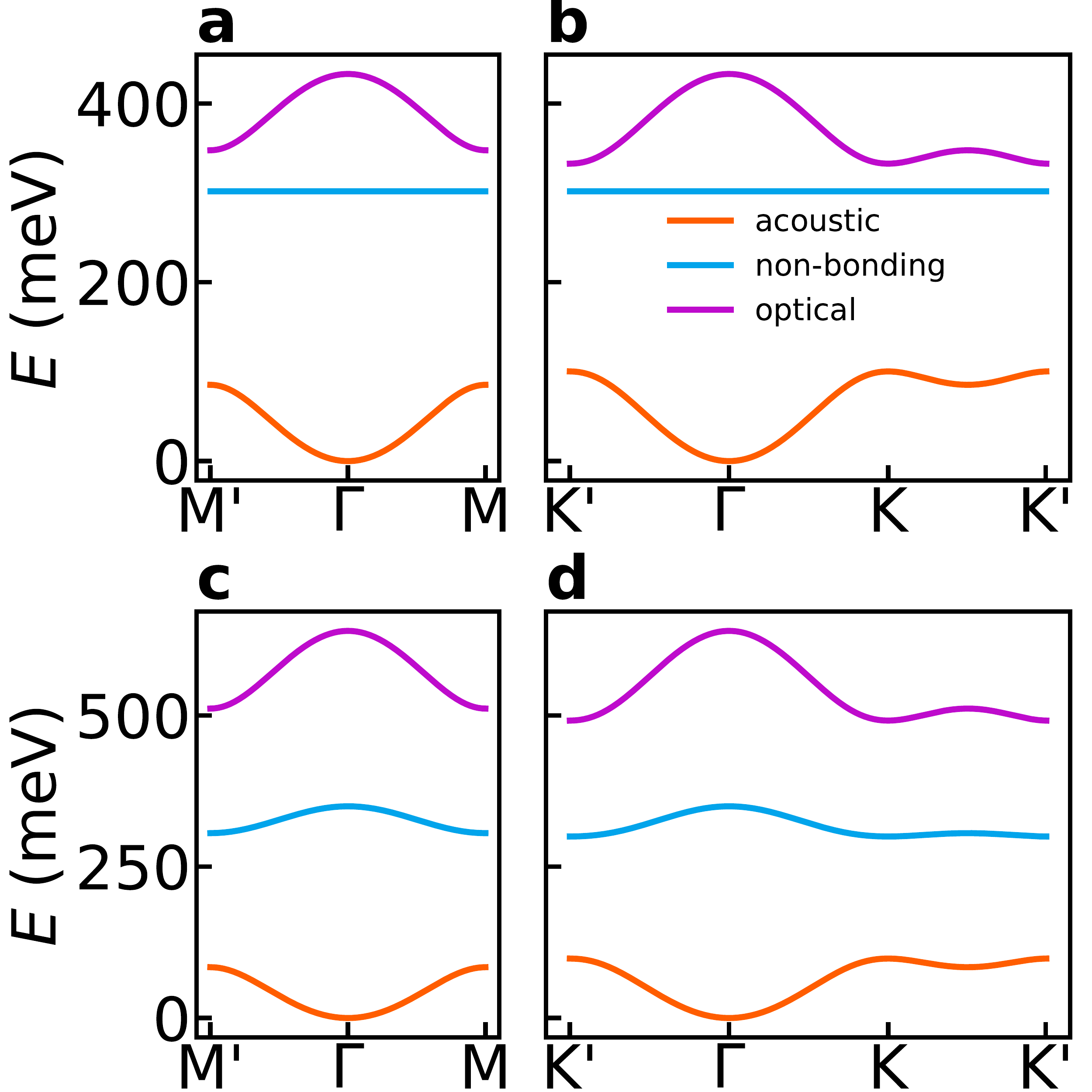}
\caption{Top panels: Magnon dispersion relations along high-symmetry lines in the triangular 
lattice's Brillouin zone, computed within the nn approximation. We take $J^{(1)}_{AB}\simeq 44$ meV, 
$J^{(1)}_{BB}\simeq 80$  meV. Bottom panels: same as above, but including nnn exchange.  
We take $J^{(1)}_{AB}\simeq 65$ meV, $J^{(1)}_{BB}\simeq 80$, $J^{(2)}_{BBs}=-4.4$ meV, 
$J^{(2)}_{BBd}=J^{(2)}_{AAd}=0$.}
\label{fig:Heisenbergnn}
\end{figure}

\subsubsection{Next nearest neighbour contribution}
In order to obtain a dispersive non-bonding band, still in the case without SOC,   it is necessary to include next nearest neighbour (nnn) exchange.
This includes the $AA$ nnn exchange  $J^{(2)}_{AA}$ that will contribute to the acoustic and optical bands, 
as well as the $BB$  exchange couplings, including the {\em same layer} $J^{(2)}_{B_1B_1}=J^{(2)}_{B_2B_2}\equiv J^{(2)}_{BBs}$  and the {\em different layer} $J^{(2)}_{B_1B_2}\equiv J^{(2)}_{BBd}$.
Unlike the nn, the nnn gives diagonal contributions both for the Ising and flip-flop exchange in the $AA$ and $BBs$ channels.  We thus have the following contributions to the Hamiltonian matrix:

\begin{eqnarray}
h^{(2)}_{AA}(\vec{k}) &=& J^{(2)}_{AA}S_A\left[6-f_2(\vec{k})\right]\nonumber\\
h^{(2)}_{BB}(\vec{k})&=& \left\{J^{(2)}_{BBs}\left[6-f_2(\vec{k})\right]+6J^{(2)}_{BBd}\right\}S_B \nonumber \\
h^{(2)}_{12}(\vec{k})&=& -S_BJ^{(2)}_{BBd}f_2(\vec{k}) 
\end{eqnarray}
where
\begin{equation}
f_2(\vec{k})=2\left[(\cos(\phi_1)+\cos(\phi_2)+\cos(\phi_1-\phi_2)\right] .
\end{equation}
We thus see that the non-bonding band dispersion, given by eq. (\ref{anali}), now reads,
\begin{eqnarray}
E_{\rm non-bonding}=
3 J^{(1)}_{AB}S_A+ J^{(1)}_{BB}S_B + \nonumber\\
+
\left\{J^{(2)}_{BBs}\left[6-f_2(\vec{k})\right]+6J^{(2)}_{BBd}\right\}S_B+S_BJ^{(2)}_{BBd}f_2(\vec{k}) 
=\nonumber\\
3 J^{(1)}_{AB}S_A+ J^{(1)}_{BB}S_B+\nonumber \\
+\left[6(J^{(2)}_{BBd}+J^{(2)}_{BBs})
+(J^{(2)}_{BBd}-J^{(2)}_{BBs})f_2(\vec{k})\right]S_B.\nonumber\\
\end{eqnarray}
We note that the opposing sign in the amplitude of the dispersive part of the of the non-bonding part arises from the antisymmetric nature of this mode. 

Given that $f_2(\Gamma)=6$ and $f_2(K)=-3$, the fact that the maximum of the non-bonding band 
is a the $\Gamma$ point indicates that $J^{(2)}_{BBd}-J^{(2)}_{BBs}$ is a positive number.  Since we expect that, in absolute value $J^{(2)}_{BBd}$ is smaller than $J^{(2)}_{BBs}$, on account of the larger distance, this probably entails that second neighbour interactions are AF.  For the sake of simplicity we assume $J^{(2)}_{BBd}=0$ and we infer   $J^{(2)}_{BBs}= -\frac{W}{9S_B}$, where $W$ is the bandwidth of the non-bonding band in the $\Gamma-K$ direction and $9=f_2(\Gamma)-f_2(K)$.   Since $W\simeq 50$ meV, we estimate $J^{(2)}_{BBs}=-4.4$ meV.

The second neighbour exchange also affects the bandwidth of the acoustic magnon band as well as the center of mass position of the non-bonding band.   As a result, we need to readjust the values of the first neighbour exchange interactions in the model to obtain magnon bands in agreement with those computed with the itinerant fermion picture. 

\subsubsection{Magnon wave functions}
\begin{figure}
\includegraphics[width=\columnwidth]{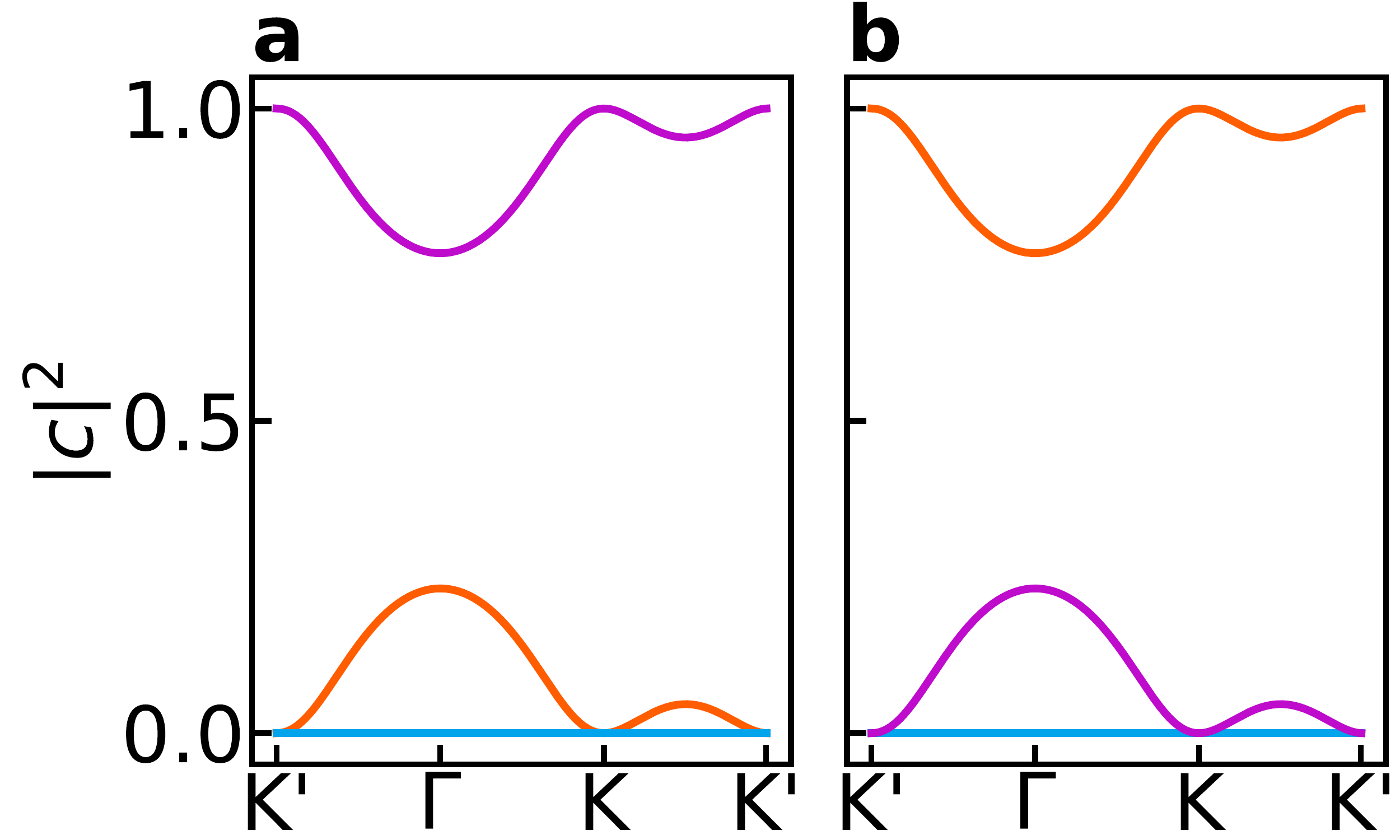}
\caption{Projection of the magnon wave functions over the $A$ site (left) and the symmetric mode
of the $B$ sublattice (right).
}
\label{fig:wavefun}
\end{figure}

In figure \ref{fig:wavefun} we plot the projection of the magnon wave functions for the three bands over the two modes that are mixed by the magnon Hamiltonian (\ref{deco}), namely, the $A$ site and the symmetric $B$ mode.
The results of the figure are computed including the nnn exchange, with the same parameters than the lower panels of Fig.  \ref{fig:Heisenbergnn}.
We see that the acoustic  
 band is predominantly located on the $B$ sublattice.  The non-bonding band has 100\%  of the weight on the anti-symmetric bond.  The $A-B$ mixing  on the acoustic and optical magnons is  minimal at the Dirac points, as expected in a honeycomb lattice. 

\subsection{Magnons with DM and anisitropy}

We now consider the two terms in the Hamiltonian,   anti-symmetric DM exchange and anisotropy, that can only be present in the spin Hamiltonian when spin-orbit interactions are present in the parent fermion Hamiltonian.

For the DM term we make the following assumptions. The first neighbor DM occurs only 
for the $AB$ dimers, and the $AB_1$ and $AB_2$ couplings are the same
We assume the second neighbor coupling is active only in the $B$ sublattice, is diagonal in 
the $B$ subspace, and identical for $B_1$ and $B_2$. 
\begin{eqnarray}
{\cal H}^{DM}_{a,a'}(\vec{k}) \equiv
 \left(
\begin{array}{ccc}
0 & d_1(\vec{k})&  d_1(\vec{k})\\
d_1^*(\vec{k}) & d_2 (\vec{k}) & 0\\
 d_1^*(\vec{k})& 0 & d_2 (\vec{k})
\end{array}
\right),
\end{eqnarray}
where  $d_1(\vec{k})= i\sqrt{S_A S_B}[f(\vec{k})-1]$ and $d_2(\vec{k})= 2 D_2 S_B 
\left[\sin(\vec{k}\cdot\vec{a}_1)+\sin(\vec{k}\cdot\vec{a}_2)+\sin(\vec{k}\cdot(\vec{a}_1-\vec{a}_2))\right]$

Both $d_1$ and $d_2$ are odd functions of the momentum, $d_{1}(\vec{k})=-(d_{1}(-\vec{k}))^*$, and
$d_{2}(\vec{k})=-d_{2}(-\vec{k})$.  However, they enter in the energy dispersion in different manners. 
The nn DM only gives a non-reciprocal dispersion along the $\Gamma-M$ direction for the acoustic band, 
leaving the non-bonding band unaffected. In contrast, the nnn DM gives a non-reciprocal dispersion along 
the $K-\Gamma,K'$ direction, and affects both the acoustic and the non-bonding bands. The nnn DM coupling 
enters linearly in the dispersion of the non-bonding band. As a result, we can use the valley splitting 
of the non-bonding band  $\delta=E_{\rm non-bonding}(K)-E_{\rm non-bonding}(K')=-4\sqrt{3}D_2 S_B$. 
Since the valley splitting is $\delta= 45$ meV, we infer $D_2=\frac{\delta}{4\sqrt{3}S_B}\simeq -5$ meV.

This model accounts for the non-reciprocal dispersion along the $K,K'$ direction and the reciprocal dispersion along the $\Gamma-M$ direction. However,  it also shifts laterally the  low energy magnon dispersion, in disagreement with the itinerant fermion results. This matter deserves further exploration.   
We also note that the linear contribution makes the energy of the acoustic band negative in a small range of momenta.  This could be easily fixed by adding the single ion anisotropy terms. 
  

\begin{figure}
\includegraphics[width=\columnwidth]{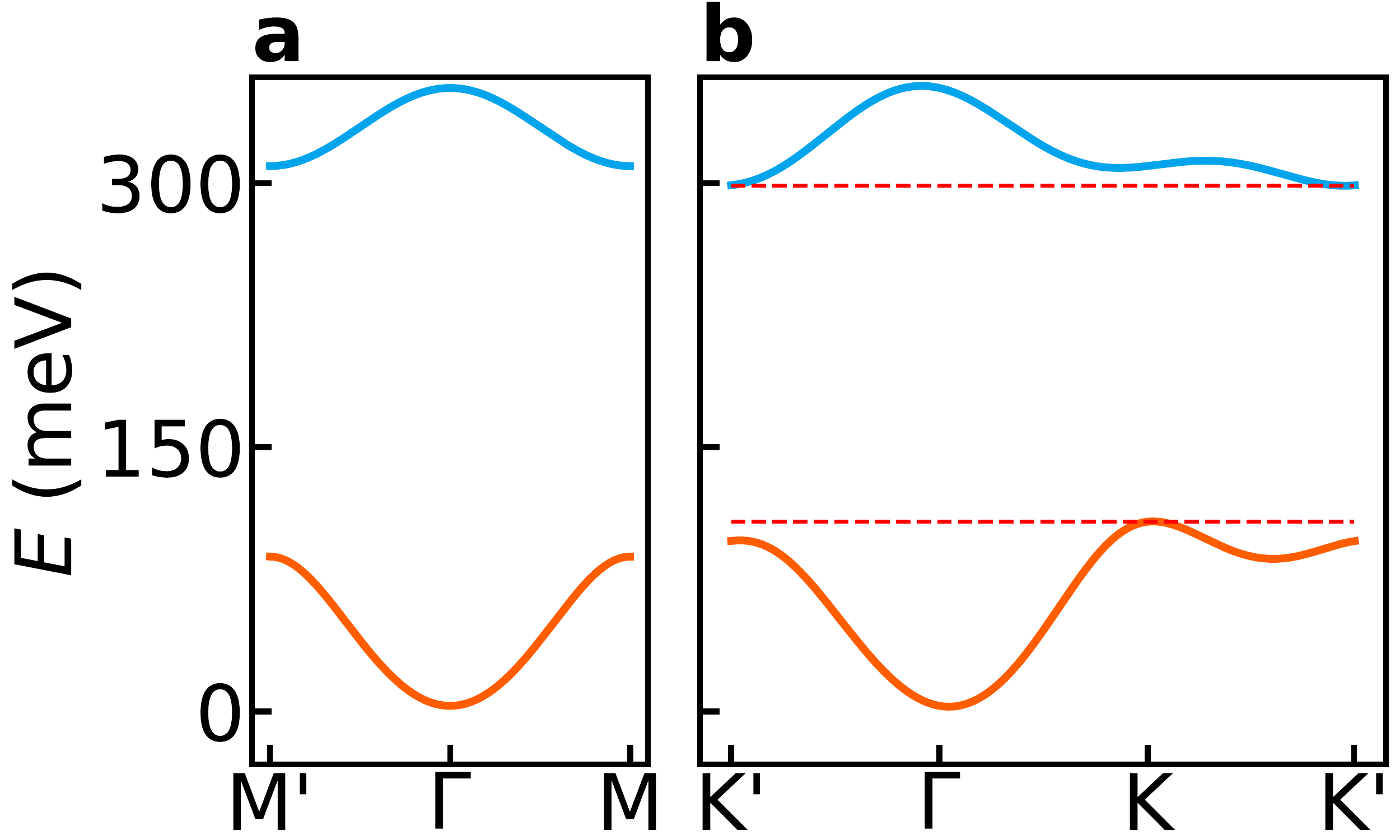}
\caption{Complete energy dispersion, including  1st and 2nd neighbour exchange and second neighbour DM interaction. We take $J^{(1)}_{AB}\simeq 65$ meV, $J^{(1)}_{BB}\simeq 80$,  $J^{(2)}_{BBs}=-4.4$ meV, $J^{(2)}_{BBd}=J^{(2)}_{AAd}=0$ and $D_2=-5$ meV. In panel b we draw red horizontal lines to emphasize the
non-reciprocity.}
\label{fig:DM}
\end{figure}



\section{Magnon wave functions from the fermionic method}

Some physical insight about the character of the magnon modes can be obtained
directly from the results of the fermionic approach. The magnon wave functions
can be obtained from the transverse spin susceptibility matrix 
$\chi^\perp_{aa'}(\omega,\vec{k})$. This object can be interpreted as the 
single-particle Green function matrix associated with the magnons. As such,
it shares eigenvectors with the magnon Hamiltonian (to which we do not have
direct access within the fermionic model). Thus, by diagonalizing 
$\chi^\perp_{aa'}(\omega,\vec{k})$ at an $(\omega,\vec{k})$ pair where
its imaginary part has a peak, we obtain the eigenvectors associated with
the magnon modes~\cite{Costa2020:CrI3,Costa2006IntermQ}. 
Of course, in metallic magnets magnons are not true eigenstates
of the system's hamiltonian, since they hybridize with the Stoner excitations.
Nevertheless, the eigenvectors of $\chi^\perp_{aa'}(\omega,\vec{k})$ still provide
useful information, as long as the magnon feature in the spectral density can 
be clearly distinguished from the background of Stoner modes.

\begin{figure}
\includegraphics[width=\columnwidth]{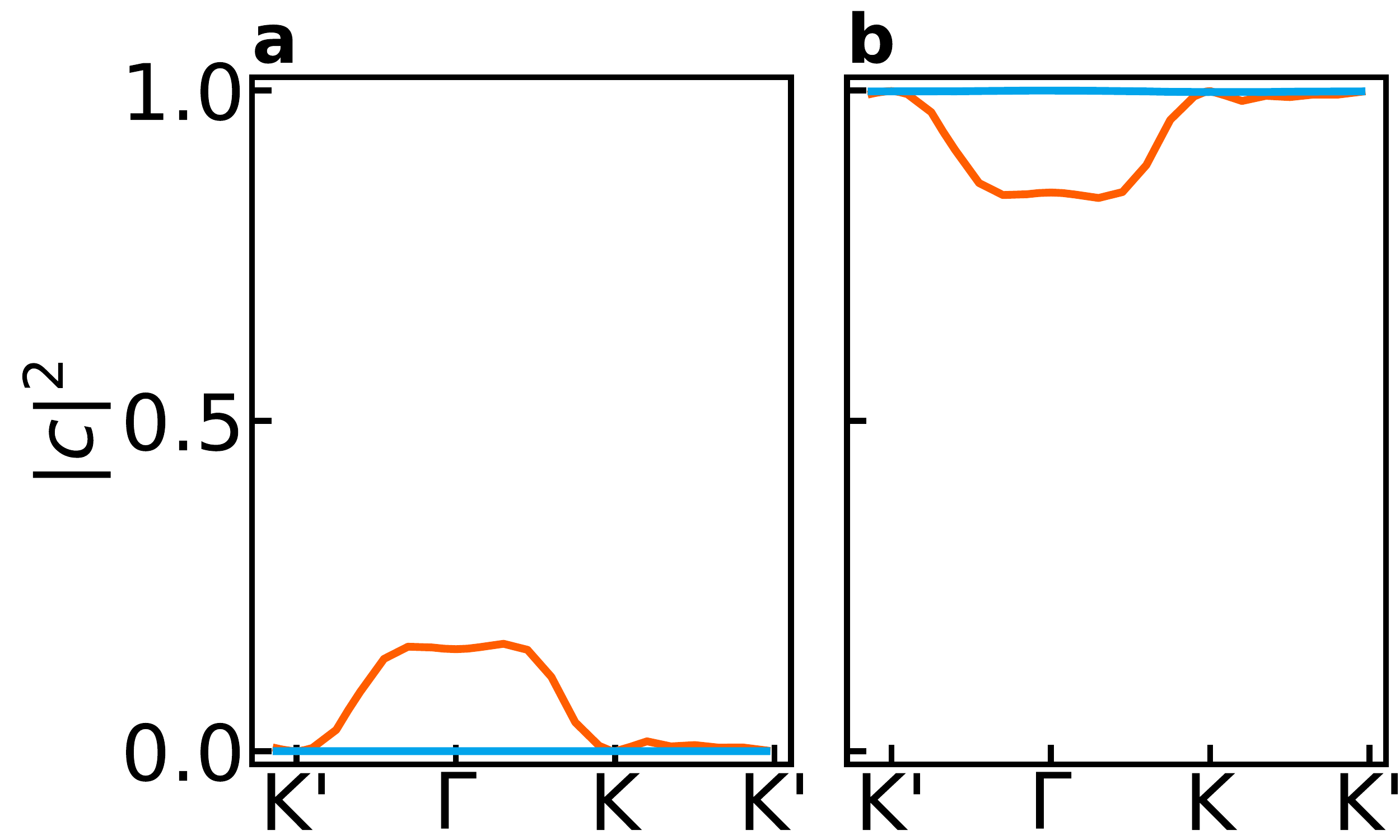}
\caption{Projections of the magnon wave functions on the $A$ site (\textbf{a}) and 
on the symmetric and antisymmetric modes of the $B$ sublattice (\textbf{b}) 
along the $\Gamma-K$ line, extracted directly from the fermionic method. 
The orange lines correspond to the projections of the acoustic mode on $A$ and
on the \textsl{symmetric} $B$ mode. The blue lines are the projections of the non-bonding 
mode on $A$ and on the \textsl{antisymmetric} $B$ mode. As discussed in the text, the third 
mode, predicted by the spin model, does not have a clearly identifiable signature in the magnon 
spectral density derived from the fermionic approach, due to strong hybridization with the 
Stoner continuum.}
\label{fig:MagnonPsi2fermion}
\end{figure}

As we have shown in the paper, \fegete\ has only two well-defined magnon modes.
By looking at the dispersion of the lowest energy mode (fig. 3b of the main paper), 
it is clear that it can be identified with the acoustic mode of the spin model of 
the previous section. Analysis of its wave function coefficients, shown in 
fig.~\ref{fig:MagnonPsi2fermion} confirms that identification.

The character of the higher-energy mode is less evident from the dispersion relations alone, 
since the spin model predicts two modes with energies higher than that of the acoustic magnons.
Comparison of the wave function coefficients obtained in the fermionic method with those obtained
from the spin model indicate that this mode corresponds to the second mode predicted by the spin 
model. This conclusion is supported not only by the fact that $|c_{A}|^2=0$ along the whole 
$\Gamma-K$ line, as seen in figs.~\ref{fig:wavefun} and~\ref{fig:MagnonPsi2fermion}, but also 
by the relative phase of the $c_{B_1}$ and $c_{B_2}$ coefficients, that both methods indicate to 
be $\pi$, independent of the magnon wave vector.

\bibliographystyle{apsrev4-1}